\documentclass{article}
\usepackage{hyperref}
\usepackage{graphicx}
\usepackage{subcaption}
\usepackage{amsmath}
\usepackage{booktabs}
\usepackage{url}
\usepackage{authblk}
\usepackage{amsmath}
\usepackage{amsthm}
\usepackage{amssymb}

\theoremstyle{plain}
\newtheorem{definition}{Definition}
\newtheorem{proposition}{Proposition}

\renewcommand{\bar}{\overline}

\newdimen\figrasterwd
\figrasterwd\textwidth

\usepackage[htt]{hyphenat}
\usepackage{dcolumn}
\newcolumntype{d}[1]{D{.}{.}{#1}}

\title{Generalizing Homophily to Simplicial Complexes}
\author[a,*]{Arnab Sarker}
\author[b]{Natalie Northrup}
\author[a,b]{Ali Jadbabaie}
\affil[a]{Institute for Data, Systems, and Society, MIT}
\affil[b]{Department of Civil and Environmental Engineering, MIT}
\affil[*]{Corresponding Author, E-mail: \texttt{arnabs@mit.edu}}
\date{June 2022}

\begin{document}

\maketitle              % typeset the title of the contribution

\begin{abstract}
  Group interactions occur frequently in social settings, yet their properties beyond pairwise relationships in network models remain unexplored.
  In this work, we study homophily, the nearly ubiquitous phenomena wherein similar individuals are more likely than random to form connections with one another, and define it on simplicial complexes, a generalization of network models that goes beyond dyadic interactions.
  While some group homophily definitions have been proposed in the literature, we provide theoretical and empirical evidence that prior definitions mostly inherit properties of homophily in pairwise interactions rather than capture the homophily of group dynamics.
  Hence, we propose a new measure, $k$-simplicial homophily, which properly identifies homophily in group dynamics.
  Across 16 empirical networks, $k$-simplicial homophily provides information uncorrelated with homophily measures on pairwise interactions.
  Moreover, we show the empirical value of $k$-simplicial homophily in identifying when metadata on nodes is useful for predicting group interactions, whereas previous measures are uninformative.
%\keywords{social network analysis, homophily, simplicial complexes}
\end{abstract}
\section{Introduction}
% \begin{itemize}
%     \item Homophily is a core feature of social networks.
%     \item Group homophily explains a lot of societal phenomonena
%     \item Current metrics for homophily are insufficient.
%     \item Here, we provide a new definition.
%     \item Provide experiments on synthetic and real data, to show why it is important to account for pairwise information.
%     \item Application to simplicial link prediction as well ?
% \end{itemize}
Group interactions fundamentally differ from interactions between pairs of individuals.
When individuals assemble in groups of size three or more, social pressure increases~\cite{asch1955opinions}, social loafing may occur~\cite{latane1979many}, and joint decisions can become polarized~\cite{brown1986social}.
However, fundamental properties of group interactions in complex networks are not yet fully explored.
As such, \emph{higher order} models, which explicitly encode group interactions with data structures such as simplicial complexes and hypergraphs, have received attention in recent literature \cite{battiston2020networks,benson2018simplicial,schaub2020random}.

In this work, we consider the principle of homophily as it pertains to group interactions.
Homophily, the well-known tendency for individuals to form social connections with those similar to themselves, is a core organizing principle of social networks \cite{lazarsfeld1954friendship,mcpherson2001birds}.
This notion is nearly ubiquitous, appearing in contexts such as marriage, friendship, information transfer, physical contact, and online social networks \cite{kossinets2006empirical,mcpherson2001birds,stehle2011high}.
In such networks, social ties are correlated with similarity in age, occupation, religion, and/or each individual's local network structure \cite{dong2017structural,mcpherson2001birds}.
Although this empirical ubiquity of homophily makes it valuable in understanding social structure, previous studies have restricted to analysis of graphs, which only encode pairwise interactions between individuals.
% The data is said to display homophily if the pairwise interactions are more likely than random to be between individuals who share the same characteristics \cite{easley2010networks}.

% Here, we introduce a new measure for homophily at the group level, analyze the presence of homophily in synthetic and empirical networks, and show the utility of homophily in the context of link prediction.

\begin{figure}[t]
  \centering
  \parbox{0.56\figrasterwd}{
    \parbox{.1\figrasterwd}{%
      \subcaptionbox{}{\includegraphics[width=\hsize]{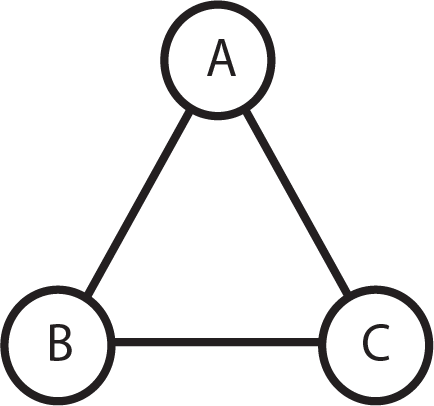}}
      \vskip 0.2em
      \subcaptionbox{}{\includegraphics[width=\hsize]{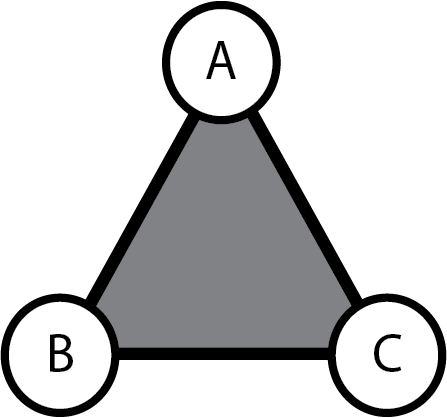}}  
    }
    \hskip 1em
    \parbox{.44\figrasterwd}{%
      \subcaptionbox{\label{fig:homophily_ex}}{\includegraphics[width=\hsize]{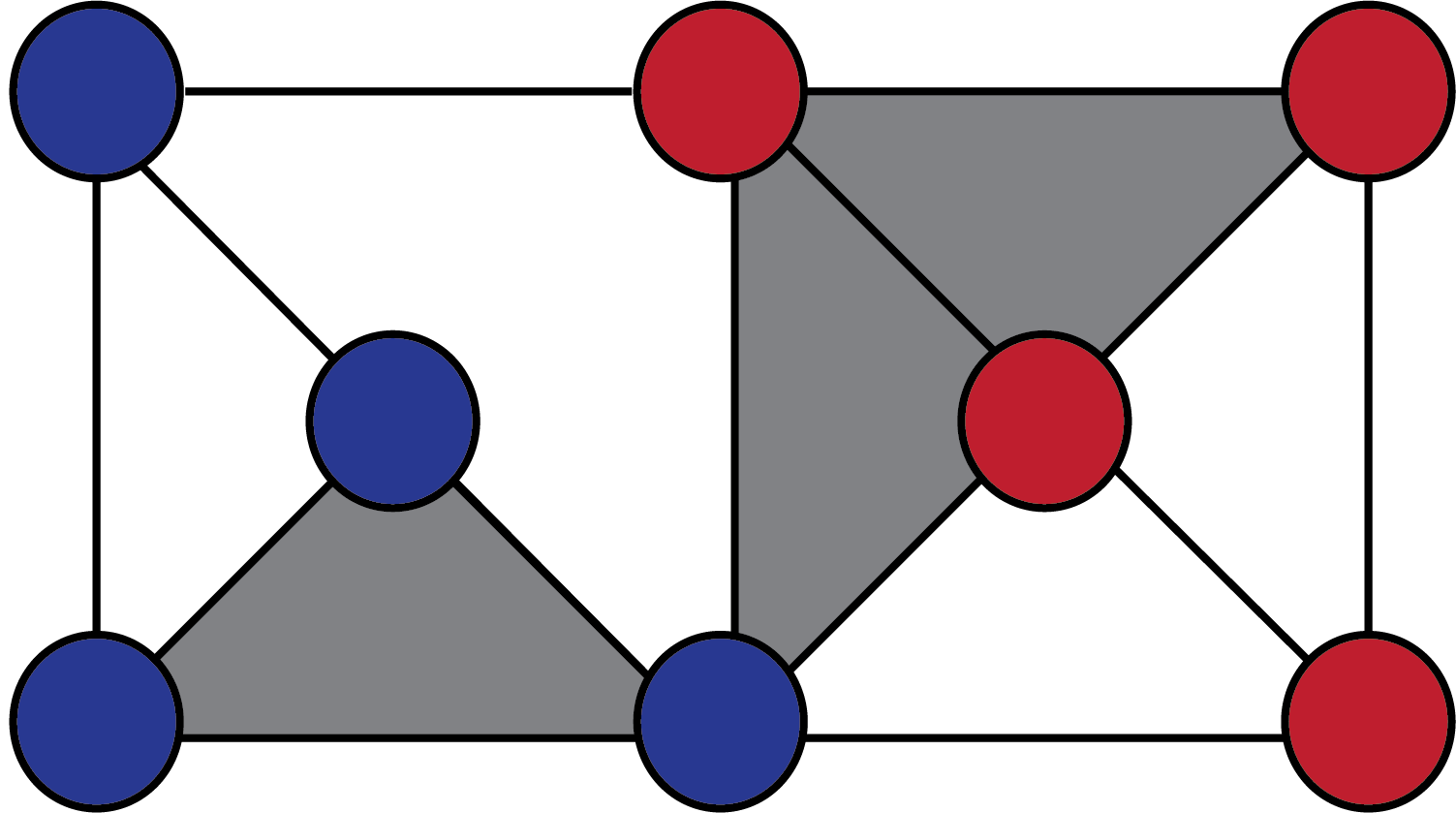}}
    }\hskip 1em
    % \parbox{.3\figrasterwd}{%
    %   \begin{align*}
    %       a^3(X) &= \frac{2}{3} \\
    %       b_h^3(X) &= \frac{1}{7}, &b_x^2(X) = \frac{4}{6} \\
    %       s_h^3(X) &= \frac{14}{3}, &s_b^2(X) = 1 
    %   \end{align*}
    % }
  }
  \caption{(a) A closed but not filled triangle, indicating only pairwise interactions (e.g., three separate two-author papers). (b) A closed and filled triangle, indicating a group interaction (e.g., a single paper by three authors). (c) Example where group homophily is inherited from edge structure. If nodes were randomly labeled, then $1/7$ filled triangles would have nodes of the same type on average (c.f. \eqref{eq:baseline_hypergraph}), suggesting the presence of homophily (2 out of 3 filled triangles have nodes of the same type, and $2/3 > 1/7$). However, the edge structure of the network is such that 4 out of  6 closed (filled or unfilled) triangles have nodes of the same type, which suggests that the homophily of filled triangles is as if they were randomly placed into the underlying edge structure.
  \vspace{-0.15in}}
\end{figure}

Our work builds on definitions of group homophily considered recently in the context of hypergraphs, a generalization of graphs that can encode interactions between arbitrarily large groups of individuals \cite{veldt2021higher}.
For a particular hypergraph with labeled nodes, prior work considers all hyperedges of fixed size $g \geq 3$, and defines homophily relative to if nodes were labeled at random, which we refer to as a \emph{node baseline}.
However, this approach can potentially inherit the dyadic, graph-based notion of homophily rather than that of group interactions. 
In other words, much of the variation of group homophily scores with a node baseline can be explained by the standard dyadic notion (Figure \ref{fig:homophily_ex}).
This observation goes beyond the provided example: in the 16 empirical datasets of this work, nearly 70\% of the variation in group homophily (for groups of size 3) using a node baseline can be explained by homophily scores defined only on edges.
Hence, we introduce a new measure, $k$-simplicial homophily, which properly isolates homophily due to group dynamics.

\emph{Contributions.} \quad
In Section \ref{s:defs}, we precisely define $k$-simplicial homophily as a formal way to account for underlying interactions in a network when establishing the presence of homophily for groups.
Rather than model the social network with a hypergraph, we use a simplicial complex which requires additional structure in the network model.
We establish theoretically that when $k$-simplicial homophily is applied to edges, we recover a standard definition of homophily on graphs, suggesting it is a natural generalization of homophily for groups.

Furthermore, contrary to the existing notions of homophily, $k$-simplicial homophily successfully isolates properties of group dynamics.
We provide theoretical evidence of this in Section \ref{s:network}, where we introduce the simplicial stochastic block model, a generative network model which allows for homophily in pairwise interactions to be decoupled from that of triadic interactions.
We show prior measures can incorrectly conclude the presence of group homophily, whereas $k$-simplicial homophily identifies group homophily if and only if the formation of triadic interactions depends on node class labels.

We then apply group homophily definitions to empirical data.\footnote{Code and experimental details can be found at \url{https://github.com/arnabsarker/SimplicialHomophily}.}
In 15 out of 16 empirical datasets, we find that homophily scores using $k$-simplicial homophily are lower than scores computed using a baseline that depends on the proportion of each class of nodes.
Moreover, in 4 of these datasets, we find anti-homophily with respect to $k$-simplicial homophily, and note that the presence of anti-homophily is justified in each dataset.
Importantly, we do not find a significant relationship between edge homophily scores and $k$-simplicial homophily scores on triangles, suggesting that $k$-simplicial homophily provides novel insights into group dynamics which are not explained by homophily in edges.

In Section \ref{s:link_prediction}, we show the utility of the new information provided by $k$-simplicial homophily in the data-driven application of higher order link prediction.
Originally proposed by Benson et al \cite{benson2018simplicial} as a benchmark problem for higher order models and algorithms, higher order link prediction involves using network information up to a certain time $t$ to predict if new group interactions will occur after time $t$.
We find that $k$-simplicial homophily indicates whether node labels are useful in the prediction task, whereas previous definitions of group homophily are uninformative in determining the utility of node labels.

\subsection{Related Work}

Group homophily has been considered for data-driven applications such as transductive learning \cite{satchidanand2015extended} and clustering \cite{kumar2020hypergraph}.
In such works, the authors use a homophily parameter as an input into a generative hypergraph model, and homophily is defined relative to a baseline distribution computed using frequencies of node class labels.
Here, we instead propose homophily measures which describe existing datasets to aid analysis of group interactions in empirical settings.

A particularly relevant work is Veldt et al \cite{veldt2021higher}, as it aims to broadly define homophily in the context of hypergraphs.
Like previous work, the baseline considered by the authors uses randomization of node labels in order to determine if hyperedges in a network are more likely than random to be among nodes of the same type.
The author's main focus in the work is understanding the complexity that arises in group homophily due to the fact that different numbers of each category of individuals can be present in a particular group.
That is, for the setting considered by the authors where nodes are given one of two labels, a hyperedge of size $k$ can have $t$ members of one group, and $k - t$ members of the other for any $0 \leq t \leq k$.
For a fixed $k$, the authors define a homophily score for each $t$, and prove impossibility results showing their homophily scores can not be strictly increasing in $t$ and can not be greater than unity for all $t \geq k/2$.
In this work, we define similar metrics for homophily which are based on simplicial complexes as opposed to hypergraphs.
We also consider a more general setting where three or more class labels are allowed, which helps to avoid  impossibility results from prior work and allows for a broader selection of data.
% \begin{itemize}
%     \item Review of Veldt et al. specifically, which defines homophily in the context of hypergraphs. 
% \end{itemize}

\section{Preliminaries}
We discuss three data structures, each of which considers a set of nodes $V$, where $|V| = n$, and a labeling function $C: V \rightarrow \{ 1, \dots, m \}$, which maps each node to one of $m \geq 2$ classes.

\emph{Graphs and Hypergraphs.} \quad
Graphs and hypergraphs are common models of interactions in complex networks \cite{battiston2020networks}.
We consider undirected graphs which consist of a set of nodes $V$ and a set of edges $E$, where each edge $e \in E$ denotes a pairwise interaction between nodes.
Hypergraphs, in contrast, have a set of hyperedges $H \subseteq 2^V$ which are unrestricted in size. 
Hence, group interactions can be encoded as elements of $H$, with no additional structure required of $H$.

\emph{Simplicial Complexes.} \quad
Simplicial complexes provide a way to encode group interactions which requires more structure than hypergraphs.
A simplicial complex is a set of simplices $X \subseteq 2^V$, where each element $x \in X$ is referred to as a $k$-simplex if it contains $k+1$ different elements of $V$.
In Figure \ref{fig:homophily_ex}, nodes would then correspond to $0$-simplices, edges to $1$-simplices, and filled triangles to $2$-simplices. 
Simplicial complexes also have the following structural property:
\[  x \in X \implies \sigma \in X, ~\forall \sigma \subseteq x \,.\]
That is, for every simplex $x$ in $X$, all subsets of $x$ must also be contained in the simplicial complex.
In Figure \ref{fig:homophily_ex} the network can be modeled as a simplicial complex because for each filled triangle, all edges associated with the triangle are in the network.
This simple assumption leads to a rich mathematical theory from algebraic topology \cite{hatcher2005algebraic}.
While we will not discuss algebraic topology at length (instead, see \cite{ghrist2014elementary,hatcher2005algebraic}), we do utilize the definition of a $k$-skeleton.
\begin{definition}[$k$-skeleton \cite{ghrist2014elementary}]
For a simplicial complex $X$, let $X^j$ denote the set of all $j$-simplices in $X$, i.e. those elements of $X$ with exactly $j + 1$ elements.
The $k$-skeleton of $X$, denoted $X^{(k)}$, is defined
\begin{equation}\label{eq:k-skeleton}
    X^{(k)} = \bigcup_{j = 0}^{k} X^j \,.
\end{equation}
\end{definition}
As we will see, the $k$-skeleton accounts for underlying interactions when defining group homophily, as it encodes all interactions of size at most $k + 1$.
For example, in Figure \ref{fig:homophily_ex}, we used the $1$-skeleton, referred to as the underlying graph, to argue that homophily in triadic interactions (filled triangles) can be inherited from homophily in closed triangles, which are defined by pairwise interactions.

\section{Defining Group Homophily}
\label{s:defs}
% \begin{itemize}
%     \item We distinguish between global and heterogeneous definitions of homophily. Global asks a binary question of the dataset: On average, is there a tendency in this data for nodes to interact with similar types?
%     \item heterogeneous definitions are closer to those considered in Veldt et al. and allow us to get a better look at the composition of groups.
% \end{itemize}
% Definitions for homophily on pairwise interactions have been proposed in the literature, and broadly fall into two categories: global definitions and heterogeneous definitions.
% Measures such as assortativity and modularity provide an aggregate notion of whether similar nodes in a particular network are more likely than random to form edges with one another \cite{newman2002assortative,newman2006modularity}, and hence constitute global measures.
% heterogeneous measures have also been proposed, which compute homophily for each class of nodes to better describe the behavior of each class in a dataset \cite{altenburger2018monophily,coleman1958relational}.

Defining homophily for groups is far more complex than for edges, as there are significantly more options for node labels to be assigned in a group of size $g \geq 3$ than there are for an edge which only contains two nodes.
To reduce this complexity, we focus on two types of homophily in this work: one based on the proportion of homogeneous groups in a network, and another which takes into account the number of individuals in a group which share a particular class label.

% Here, we consider two distinct but related approaches to determining the presence of homophily in a network.
% The first set of definitions determine whether homophily is present in a global sense, similar to assortativity, and asks if groups are more likely than random to have all of their nodes be of the same type.
% The second set of definitions focus on a heterogeneous measure, and aim to quantify group homophily for a particular class $c \in \{ 1, \dots, m\}$ as the number of members of class $c$ in the group changes.
% This second approach shares similarities with the higher order setting of Veldt et al \cite{veldt2021higher}, in which we expand on previous work by introducing a different baseline measure for homophily which accounts for underlying lower order interactions of the network.

\subsection{Homophily of Homogeneous Groups}
\label{ss:global}
% The definitions considered here require (1) an affinity score equal to the observed proportion of homogeneous groups (groups for which all nodes have the same label), and (2) a baseline score equal to a proportion of homogeneous groups in expectation.
% We consider two measures which differ in the second component.
In what follows, we refer to a group as homogeneous if all nodes share the same class. 
We use $g$ to refer to group size in a hypergraph, and $k$ to refer to $k$-simplices in a simplicial complex, which have size $g = k + 1$.
For an arbitrary hypergraph $H$, let $H^g$ represent the hyperedges of size $g$ and $H^g_h \subseteq H^g$ represent the homogeneous hyperedges of size $g$.
The affinity score is then defined
\begin{equation} \label{eq:affinity_hypergraph}
a^g(H) = \left\vert H^g_h \right\vert ~/~ \left\vert H^g\right\vert  \,. 
\end{equation}
The following random baseline formalizes notions of higher order homophily from previous literature \cite{satchidanand2015extended,kumar2020hypergraph,veldt2021higher}, and can be applied to arbitrary hypergraphs:
\begin{equation} \label{eq:baseline_hypergraph}
b_h^g(H) = \sum_{c = 1}^m \binom{n_c}{g}/\binom{n}{g} \,,
\end{equation}
where $n_c$ represents the number of individuals in class $c$, so
$b_h^g(H)$ represents probability that a group of size $g$ in $H$ with random node labels is homogeneous.
\begin{definition}[Hypergraph Homophily Score~\cite{veldt2021higher}]
The hypergraph homophily score $s_h^g(H)$ is defined
\begin{equation} \label{eq:hypergraph_homophily}
s_h^g(H) = a^g(H) ~/~ b_h^g(H) \,.
\end{equation}
\end{definition}
The score $s_h^g(H) $ indicates the presence of homophily if $s_h^g(H)  > 1$, or anti-homophily if $s_h^g(H) < 1$.
The score also coincides with a traditional metric of graph homophily when $g = 2$, which we refer to as the \emph{graph homophily score}.

The second random baseline applies only to simplicial complexes.
Let $\bar{X^{(k-1), k}}$ represent the possible $k$-simplices that may occur in $X$.\footnote{Formally, given $X^{(k-1)}$ is the $(k-1)$-skeleton of $X$, $\bar{X^{(k-1), k}}$ represents the maximal set of $k$-simplices which could be added to $X^{(k-1)}$ while preserving that $X^{(k-1)} \cup \bar{X^{(k-1), k}}$ is a simplicial complex.}
The baseline is then
\begin{equation}
b_x^k(X) = a^{k+1}(\bar{X^{(k-1), k}})  \,. \label{eq:baseline_simplicial}
\end{equation}
Intuitively, $b_x^k(X)$ is the probability that a randomly placed $k$-simplex into the $(k-1)$-skeleton of $X$ is homogeneous.
The corresponding homophily score is:
\begin{definition}[$k$-Simplicial Homophily Score]
The $k$-simplicial homophily score $s_x^k(X)$ is
\begin{equation} \label{eq:simplicial_homophily}
s_x^k(X) = a^{k+1}(X)~/~b_x^k(X) \,.
\end{equation}
\end{definition}
The primary difference between $k$-simplicial homophily and hypergraph homophily lies in the definition of the baseline score. 
In hypergraph homophily, the baseline depends only on the composition of nodes, whereas for $k$-simplicial homophily, the $(k-1)$-skeleton accounts for the underlying interactions.

\noindent \textbf{Example.} \quad In Figure \ref{fig:homophily_ex}, consider the case $k = 2$, such that we are focused on triangles. 
Then, $X^{(k-1)}$ represents the underlying graph, and $\bar{X^{(k-1), k}}$ represents closed triangles in the underlying graph. Because 4 out of 6 closed triangles are homogeneous, $b_x^2(X) = 4/6$, and similarly because 2 out of 3 filled triangles are homogeneous, $a^3(X) = 2/3$. Therefore, $s_x^2(X) = 1$.

Notably, simplicial homophily and hypergraph homophily coincide when edges are the focus, as both generalize the standard definition of homophily in edges.
\begin{proposition} \label{prop:equivalence}
    Let $G = (V, E)$ represent an undirected graph, and let $C: V \rightarrow \{1, \dots, m\}$ represent a labeling of nodes into $m$ classes.
    Then, the graph homophily score and the $k$-simplicial homophily score on edges coincide.
\end{proposition}
% \begin{proof}
% Because the $k$-simplicial homophily score is defined on edges, and the graph homophily score is equivalent to the hypergraph homophily score with $g = 2$, that the affinity scores coincide (that is, $a^2(X(G)) = a^2(E)$).
% What remains is to establish that the random baseline scores are identical, i.e. $b_x^1(X(G)) = b_h^2(E)$.
% We see that for $k = 1$, $\left\vert \bar{X(G)^{(k-1), k}} \right\vert = \binom{n}{2}$, as the $0$-skeleton of $X(G)$ is just the set of nodes in $G$, and there are $\binom{n}{2}$ possible edges that can be formed between them while trivially maintaining the inclusion property of simplicial complexes. 
% Moreover, the number of these edges which are homogeneous, is equal to $\sum_{c = 1}^m \binom{n_c}{2}$, as there are $\binom{n_c}{2}$ possible homogeneous edges in each class.
% Hence, we see
% \[b_x^1(X(G)) = \sum_{c = 1}^m \binom{n_c}{2} / \binom{n}{2} = b_h^2(E) \,,\]
% proving the claim.
% \end{proof}
The proof of the claim follows directly from the equivalence of \eqref{eq:baseline_hypergraph} and \eqref{eq:baseline_simplicial} when applied to edges.
Proposition \ref{prop:equivalence} shows that $k$-simplicial homophily is actually a natural extension of graph-based notions of homophily \cite{easley2010networks,newman2002assortative}.
However, the two approaches to homophily differ when group size increases beyond 2, as edges in a simplicial complex can impose structure on triads.

\subsection{Homophily in Heterogeneous Groups}
\label{ss:class-specific}
While the scores of the previous section conveniently summarize homophily into a single value, they cannot handle heterogeneity in node labels for a group.
To handle this distinction, we focus on type-$t$ interactions as defined in \cite{veldt2021higher}. 
For a class $c$ and group size $g$, a type-$t$ interaction is an interaction with exactly $t$ members from class $c$.
The type-$t$ affinity score for class $c$ is defined \cite{veldt2021higher}:
\begin{equation}\label{eq:class_affinity}
    a_{c}^g (t; H) = t \times \left\vert H^{t, g}_{c} \right\vert ~/~ \sum_{i = 1}^{g} i \times \left\vert H^{i, g}_{h, c} \right\vert \,,
\end{equation}
where $H^{i, g}_{h, c}$ is the set of type-$i$ hyperedges for class $c$.
The random baselines for the heterogeneous scores are then
\begin{equation} \label{eq:hypergraph_class_baseline}
    b_h^g(t; H) = \frac{\binom{n_c - 1}{t - 1} \times \binom{n - n_c}{g - t}}{ \binom{n - 1}{g - t}} \,, \quad \text{and} \quad b_x^k(t; X) = a_{c}^{k+1} (t; \bar{X^{(k-1), k}}) \,,
\end{equation}
where the former can be shown to be the expectation of $a_{c}^g (t; H)$ when node labels are assigned randomly, and the latter generalizes the randomization scheme of Section \ref{ss:global}. 
The heterogenous homophily scores can then be defined as follows.
\begin{definition}[Heterogeneous Homophily Scores]
For a hypergraph $H$, group size $g$, and class $c$, the \emph{heterogenous hypergraph homophily score} is \cite{veldt2021higher}
\begin{equation} \label{eq:hypergraph_class_homophily}
s_{h, c}^g(t; H) = a_{c}^g (t; H) ~/~ b_h^g(t; H) \,.
\end{equation}
For a simplicial complex $X$, the \emph{heterogeneous $k$-simplicial homophily score} is
\begin{equation} \label{eq:simplicial_class_homophily}
s_{x, c}^{k}(t; H) = a_{c}^{k+1} (t; X) ~/~ b_x^k(t; H) \,.
\end{equation}
\end{definition}
These definitions provide additional granularity when understanding homophily.
However, we note that they do suffer from impossibility results which limit how scores can behave with respect to the group type parameter $t$ \cite{veldt2021higher}.

% In the following section, we apply the aforementioned definitions to synthetic and empirical data.
% We begin with the synthetic model to validate the interpretation of the homophily data and to show that $k$-simplicial homophily does not inherit the tendency for lower order interactions to be homogeneous, whereas hypergraph homophily scores often can.

\section{Homophily in Network Data}
\label{s:network}

\begin{figure}[t]
    \centering
    \begin{subfigure}[b]{0.48\textwidth}
         \centering
         \includegraphics[width=\textwidth]{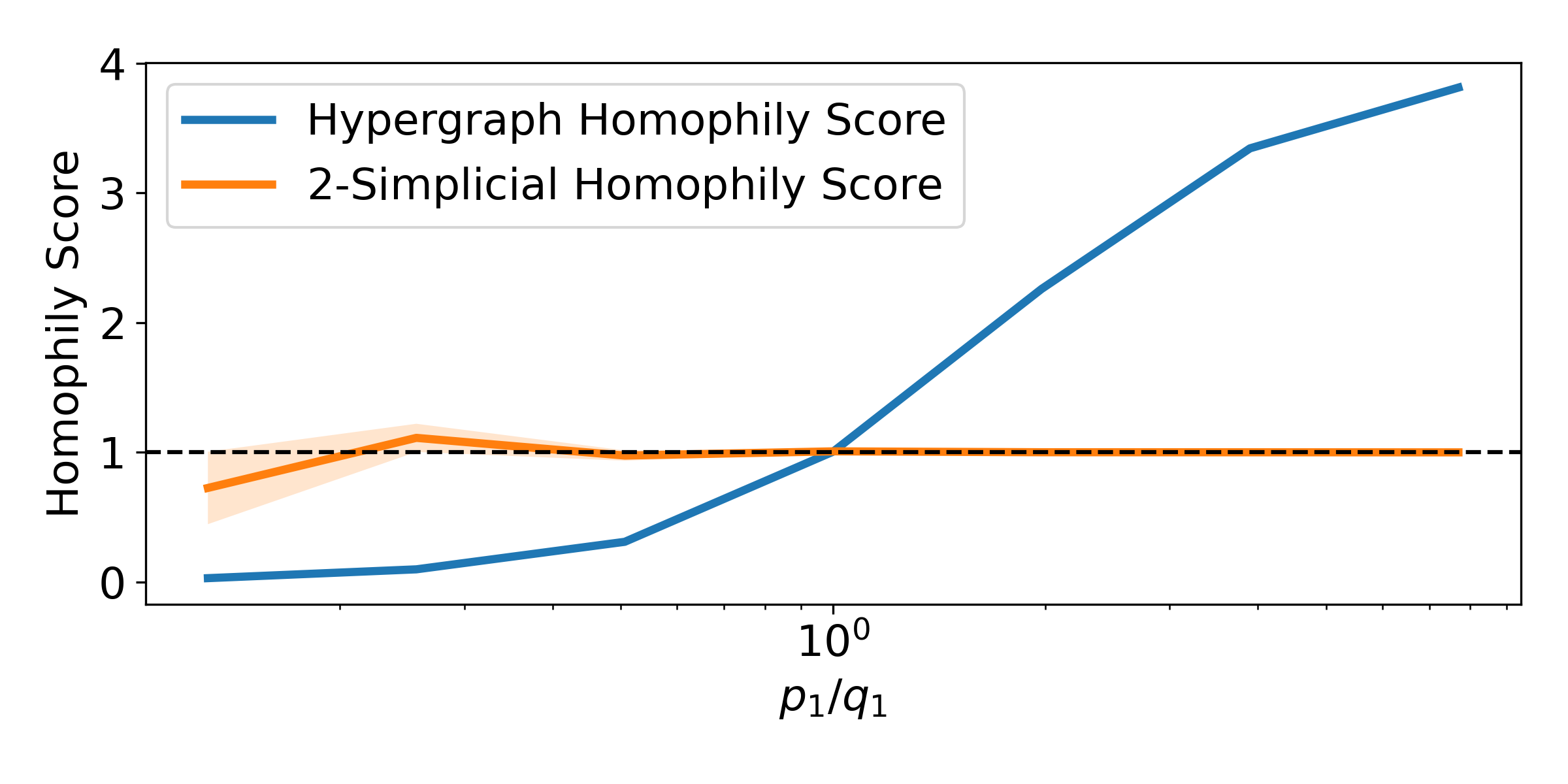}
     \end{subfigure}
     \begin{subfigure}[b]{0.48\textwidth}
         \centering
         \includegraphics[width=\textwidth]{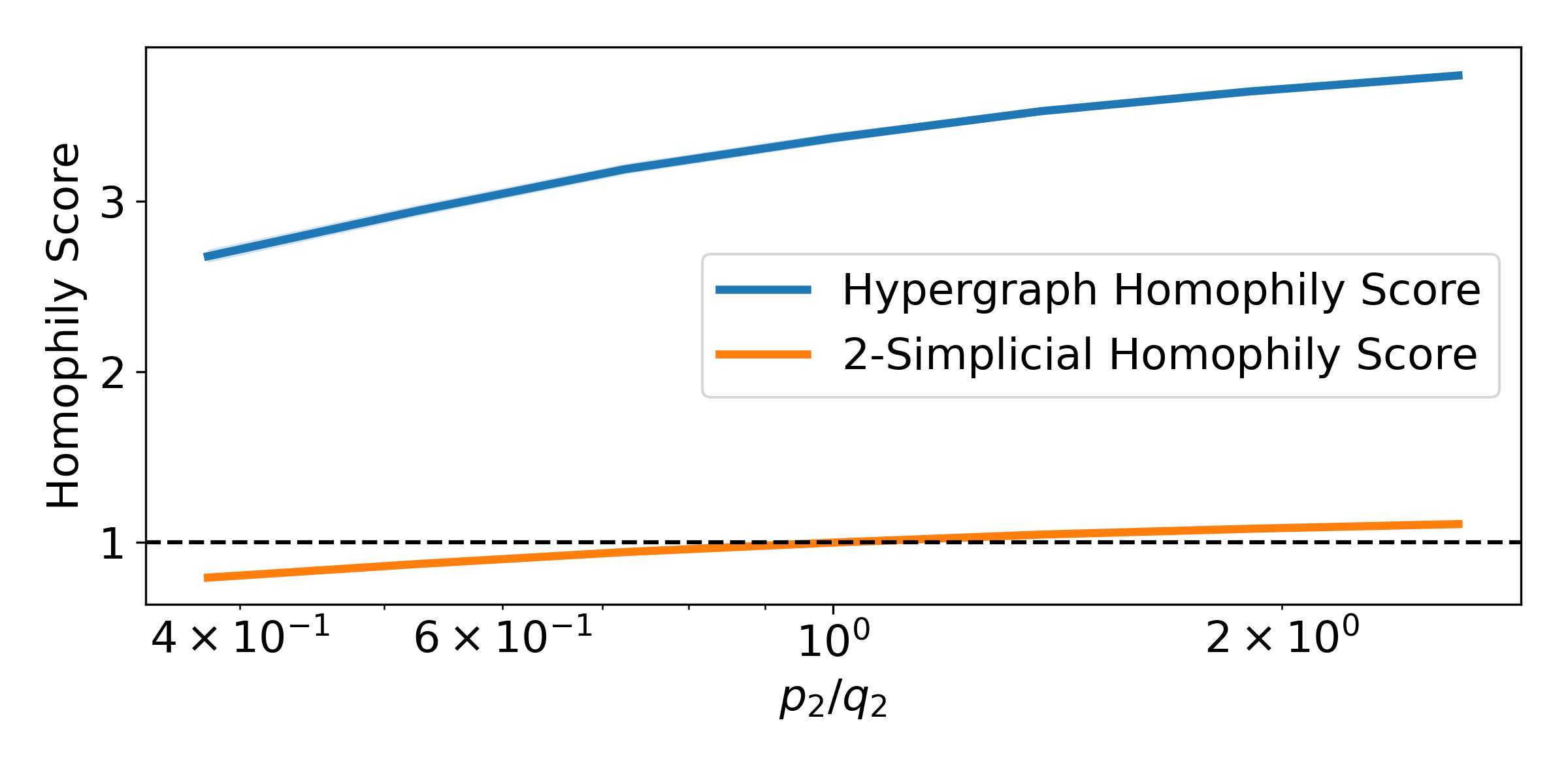}
     \end{subfigure}
    \caption{Triadic homophily in the simplicial stochastic block model.
    Error bars represent 95\% confidence intervals.
    (Left) Varying $p_1 / q_1$ with $p_2 = q_2 = 0.5$.
    The hypergraph homophily score defined on triangles inherits the homophily due to edges, whereas the $2$-simplicial homophily is near 1.
    (Right) Varying $p_2 / q_2$ with $p_1 = 4 q_1$. The $k$-simplicial homophily score is larger than 1 if and only if  $p_2 / q_2 > 1$, and hence correctly captures homophilous group dynamics. 
    In contrast, since $p_1 > q_1$, hypergraph homophily scores are consistently inflated.}
    \label{fig:ssbm_results}
\end{figure}

\begin{table*}[t]
    \centering
    \addtolength{\leftskip} {-2cm} % increase (absolute) value if needed
    \addtolength{\rightskip}{-2cm}
    \caption{Summary statistics for the 16 datasets used for homophily comparisons, 9 of which are also used for link prediction experiments.}
    \label{tab:network_data}
    \begin{tabular}{lrrrrr}
\toprule
     Dataset & nodes & classes & edges & triangles~& time steps \\
\midrule
                             \texttt{cont-village}~\cite{ozella2021using} &       46 &       5  &        329  &        610 &  \rule[.5ex]{1em}{0.5pt} \\
                              \texttt{cont-hospital}~\cite{genois2018can} &       81 &       5 &                       1,381 &      6,268 &                   12,605 \\
                         \texttt{cont-workplace-13}~\cite{genois2018can}  &      100 &       5 &    3,915 &    80,173 &                   20,129 \\
                         \texttt{email-Enron}~\cite{benson2018simplicial} &      148 &       2 &      1,344 &     1,159 &  \rule[.5ex]{1em}{0.5pt} \\
                          \texttt{cont-workplace-15}~\cite{genois2018can} &      232 &      12 &    16,725 &  329,056 &                   21,536 \\
                        \texttt{cont-primary-school}~\cite{genois2018can} &      241 &      11 &     8,317 & 5,139 &                    3,124 \\
 \texttt{bills-senate}~\cite{fowler2006connecting,fowler2006legislative} &      297 &       4 &   10,555 &      11,460 &                    4,975 \\
                           \texttt{cont-high-school}~\cite{genois2018can} &      326 &       9 &      5,818 &    2,370 &                    8,938 \\
 \texttt{bills-house}~\cite{fowler2006connecting,fowler2006legislative}  &    1,495 &       3 &       29,959 &     16,884 &                    4,871 \\
                           \texttt{hosp-DAWN}~\cite{benson2018simplicial} &    2,558 &     364 &  124,155 &  1,081,440 &                        8 \\
                       \texttt{soc-youtube}~\cite{mislove2007measurement} &   10,513 &      10 &    85,134  &     24,903 &  \rule[.5ex]{1em}{0.5pt} \\
                        \texttt{soc-flickr}~\cite{mislove2007measurement} &   54,104 &      10 &  1,231,068 &  2,692,349 &  \rule[.5ex]{1em}{0.5pt} \\
                             \texttt{coauth-dblp}~\cite{agarwal2016women} &  105,256 &       2 &          316,631 &  384,549 &                       55 \\ \texttt{clicks-trivago}~\cite{benson2018simplicial} &  172,737 &     160 &    176,194 &      116,264 &  \rule[.5ex]{1em}{0.5pt} \\
                   \texttt{soc-livejournal}~\cite{mislove2007measurement} &  259,865 &      10 &      329,954 &      176,547 &  \rule[.5ex]{1em}{0.5pt} \\
                         \texttt{soc-orkut}~\cite{mislove2007measurement} &  399,314 &      10 &    1,120,880 &  17,339 &  \rule[.5ex]{1em}{0.5pt} \\
\bottomrule
\end{tabular}
\end{table*}
\paragraph{Synthetic Networks} \quad
%\label{ss:synthetic}
In order to show the difference in homophily definitions, we build upon recent models of random simplicial complexes to introduce the \emph{simplicial stochastic block model}, a straightforward generalization of the $\Delta$-ensemble of Kahle \cite{kahle2014topology}.
The generative model has the following inputs:
\begin{itemize}
    \item $n_1, \dots, n_m$, the number of nodes in each class for the model.
    \item $p_1$ and $q_1$, the probability of an edge (1-simplex) forming between nodes in the same or different communities, respectively.
    \item $p_2$, the probability that a closed triangle consisting of nodes in the same community becomes filled as a $2$-simplex.
    \item $q_2$, the probability that a closed triangle consisting of nodes in different communities becomes filled.
\end{itemize}
Each random simplicial complex is then built using a generative process.
First, edges form between communities with probabilities $p_1$ and $q_1$ as noted above, creating a graph $G$.
Then, for each closed triangle in $G$, the triangle becomes filled with probability $p_2$ if all nodes in the closed triangle are of the same community, or with probability $q_2$ otherwise.

This model can control the presence of homogeneous edges and homogeneous triangles in the network while maintaining the structural requirement of a simplicial complex.
$p_1$ and $q_1$ determine whether there is homophily in pairwise interactions, and
$p_2$ and $q_2$ dictate how much homophily occurs in the filled triangles beyond that of the underlying pairwise interactions.

We provide two sets of experimental results on the simplicial stochastic block model, each using two classes of nodes and community sizes of 1,000 for each class.
In the left of Figure \ref{fig:ssbm_results}, we set $p_2 = q_2$ which indicates that by construction group formation is not influenced by class labels.
$k$-simplicial homophily detects that $p_2 = q_2$ and reports a value close to 1, whereas the value reported by hypergraph homophily depends on the parameters $p_1$ and $q_1$.
The hypergraph homophily score in this case is above 1 if and only if $p_1/q_1 > 1$, indicating that hypergraph homophily defined on \emph{triangles} inherits the properties of \emph{edge} homophily prescribed in the model.
In contrast, the right figure illustrates that $k$-simplicial homophily can effectively identify whether group dynamics are homophilous.
We let $p_1 > q_1$ and vary the ratio $p_2 / q_2$.
The $k$-simplicial homophily score on triangles is above 1 if and only if $p_2/q_2$ is above 1, whereas the hypergraph score is consistently above 1 because $p_1 > q_1$.
Because the hypergraph score is influenced by both $p_1/q_1$ and $p_2 / q_2$, it can not decouple their effects.

\begin{figure}[t]
    \centering
    \includegraphics[width=0.5\textwidth]{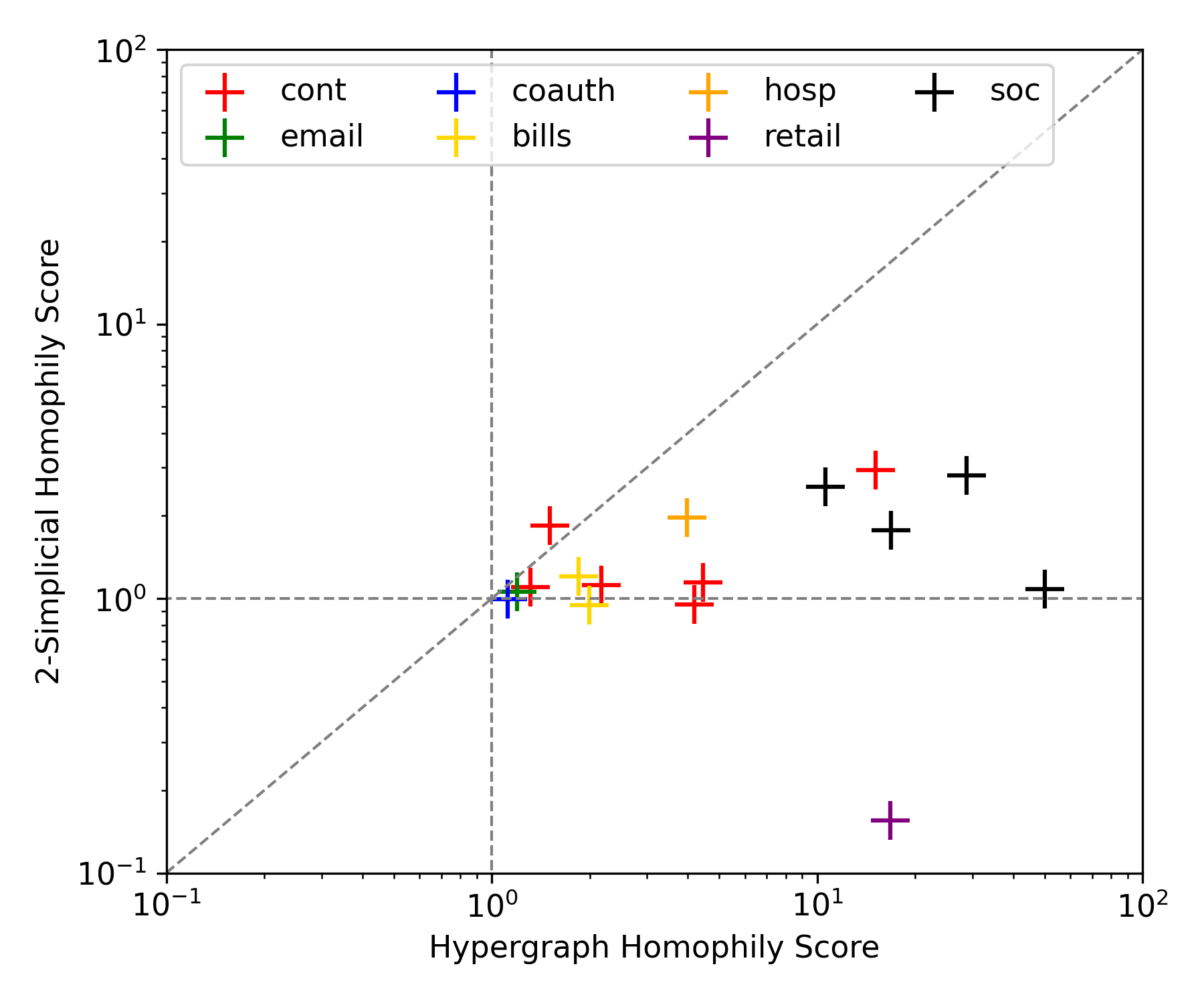}
    \caption{Scatterplot of global homophily scores with a hypergraph baseline compared to a simplicial complex baseline.
    In 15 out of 16 datasets, the $2$-simplicial homophily score is lower than the hypergraph homophily score, as the hypergraph homophily score inherits properties from edges.}
    \label{fig:global_comp}
    \vspace{-0.1in}
\end{figure}

\paragraph{Empirical Networks} \quad 
%\label{ss:empirical}
To understand the effect of different homophily definitions in empirical networks, we apply the definitions to the 16 publicly available datasets described in Table \ref{tab:network_data}.
With the empirical networks, we are able to quantify the difference between the $k$-simplicial homophily score and the hypergraph homophily score for triadic interactions. 
Using the definitions from Section \ref{ss:global}, we compute the homogeneous homophily scores for all 16 datasets and display them in Figure \ref{fig:global_comp}. 
For all but one dataset (\texttt{contact-hospital}), the hypergraph homophily score is higher than the $k$-simplicial homophily score, consistent with synthetic experiments where $p_1 > q_1$.
The result is particularly strong for \texttt{retail-trivago}, \texttt{cont-high-school}, \texttt{bills-house}, and \texttt{coauth-dblp}, for which $k$-simplicial homophily suggests anti-homophily in group formation.
In \texttt{retail-trivago}, which has the strongest tendency for anti-homophily, we posit that travelers headed to a specific destination might look at two hotels to compare cost and amenities, but if a traveler is browsing more than three hotels, they are likely taking a longer trip or have more flexibility for their search. 
For the remaining three datasets, the tendency for anti-homophily is much smaller, but can still be explained by a desire for diversity in larger group sizes.

\begin{figure}[t]
    \centering
    \includegraphics[width=0.8\textwidth]{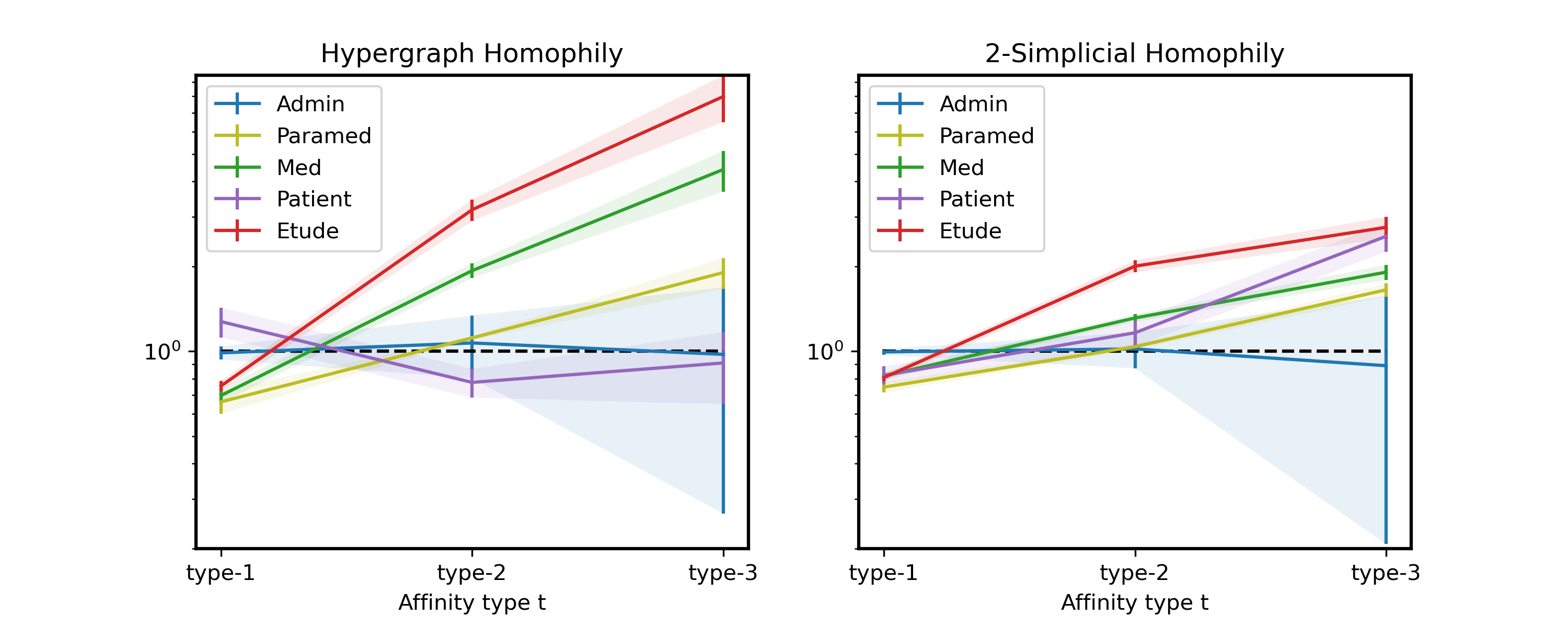}
    \caption{Homophily Scores with Heterogenous Group Compostion for \texttt{cont-hospital}. 
    Error bars represent 95\% confidence intervals. 
    We find that the simplicial complex baseline often results in less extreme values of homophily for the majority of classes in the data and all values of $t$, suggesting that pairwise interactions account for much of what is observed in hypergraph homophily. }
    \label{fig:classbased_comp}
\end{figure}

In the context of heterogeneous homophily definitions, it appears that pairwise interactions explain much of hypergraph homophily, as suggested in Figure \ref{fig:classbased_comp}.
For nearly all classes and each value of $t$, the observed metric is closer to the random baseline of $k$-simplicial homophily than that of hypergraph homophily.
That is, the baseline in $k$-simplicial homophily tends to  ``flatten'' the homophily scores as a function of type $t$.
This intuition is confirmed with homogeneous homophily scores. When using the graph homophily score to predict hypergraph homophily, we find that a simple linear model results in an $R^2$ value of $0.698$ ($p < 0.001$), with a positive coefficient that further indicates that edge homophily positively influences hypergraph homophily.
In contrast, the same analysis using graph homophily to explain $k$-simplicial homophily on triads results in an $R^2$ value of $0.167$ ($p = 0.117$), suggesting that $k$-simplicial homophily offers distinct insights on group dynamics.
In particular, we show that this distinct information is particularly useful in the task of higher order link prediction.

\section{Homophily and Higher Order Link Prediction}
\label{s:link_prediction}

\begin{table*}[t]
    \centering
    \addtolength{\leftskip} {-2cm} % increase (absolute) value if needed
    \addtolength{\rightskip}{-2cm}
    \caption{Group Formation Prediction Performance. Prediction performance is measured using the AUC-PR and presented relative to a random baseline. Bolded entries indicate a statistically significant larger performance metric via a  bootstrapping procedure which also produces confidence intervals.
    Table is sorted by $2$-simplicial homophily score of the training set, and shows that extreme $2$-simplicial homophily scores indicate when node labels are useful.}
    \label{tab:linkprediction_results}
    \begin{tabular}{lrrrr}
\toprule
 &  \multicolumn{2}{c}{Prediction Performance} & \multicolumn{2}{c}{Homophily Score} \\ \cmidrule(lr){2-3} \cmidrule(lr){4-5}
                     Dataset & Without Labels & With Labels & $2$-Simplicial $\downarrow$ &  Hypergraph \\
\midrule
         \texttt{bills-house} &           1.12 \scriptsize{($\pm$0.016)} &  \textbf{1.18} \scriptsize{($\pm$0.031)} &                            0.92 &                            2.01 \\
         \texttt{coauth-dblp} &           1.25 \scriptsize{($\pm$0.029)} &  \textbf{1.42} \scriptsize{($\pm$0.037)} &                            0.99 &                            1.12 \\
   \texttt{cont-workplace-13} &  \textbf{2.36} \scriptsize{($\pm$0.019)} &           2.22 \scriptsize{($\pm$0.016)} &                            1.05 &                            1.30 \\
        \texttt{bills-senate} &  \textbf{4.74} \scriptsize{($\pm$0.257)} &           3.38 \scriptsize{($\pm$0.161)} &                            1.16 &                            1.76 \\
   \texttt{cont-workplace-15} &           1.16 \scriptsize{($\pm$0.001)} &  1.16 \scriptsize{($\pm$0.001)} &                            1.17 &                            3.87 \\
 \texttt{cont-primary-school} &  1.08 \scriptsize{($\pm$0.000)} &           1.08 \scriptsize{($\pm$0.000)} &                            1.34 &                            2.03 \\
       \texttt{cont-hospital} &           3.38 \scriptsize{($\pm$0.028)} &  \textbf{4.46} \scriptsize{($\pm$0.038)} &                            1.79 &                            1.56 \\
           \texttt{hosp-DAWN} &           4.48 \scriptsize{($\pm$0.001)} &  \textbf{4.50} \scriptsize{($\pm$0.001)} &                            2.36 &                            6.82 \\
    \texttt{cont-high-school} &           1.48 \scriptsize{($\pm$0.001)} &  \textbf{1.55} \scriptsize{($\pm$0.001)} &                            2.84 &                            8.05 \\
\bottomrule
\end{tabular}
\vspace{-0.048in}
\end{table*}

%% Make link to homophily more clear
Higher order link prediction has been introduced as a ``benchmark problem to assess models and algorithms that predict higher-order structure''~\cite{benson2018simplicial}.
One is given a partial time series of network data up to a time $t$, and then is asked to predict if a closed but not filled triangle will become filled the after time $t$.
In the prediction task, we learn two separate logistic regression models on the first 50\% of simplices observed in the data, and test the logistic regression model on the remaining 50\% of data.
The first model (``Without Labels'') serves as a baseline and uses the local features described in Benson et al \cite{benson2018simplicial} to predict the binary outcome of whether a particular closed but not filled triangle will become filled.
The features of this regression include the frequency with which each tie occurs between each pair of nodes in the closed triangle, the degree of each node (in the traditional graph sense and weighted by the number of simplices each node is in), the number of common neighbors between the nodes, and logarithmic rescalings of all of these factors.
The second model (``With Labels'') uses the features of the first model and an additional binary indicator feature which is 1 if and only if all nodes in the closed triangle are the same type.

The results of the logistic regression are presented in Table \ref{tab:linkprediction_results}.
We evaluate performance of different features using the area under the precision-recall curve (AUC-PR) and report the score relative to a random baseline, as has been done in the literature \cite{benson2018simplicial}.
The table is sorted by the $2$-simplicial homophily score computed on the training set of data, i.e. the first 50\% of simplices which are used to train the logistic regression model. 
We find that for extreme values of the $2$-simplicial homophily score, indicating either homophily or anti-homophily, that the prediction performance increases when homogeneous node labels are used as a regressor.
Specifically, the two lowest $2$-simplicial homophily scores and the three highest $2$-simplicial homophily scores are for datasets where node labels increase predictive performance, whereas the four datasets with moderate scores see no change or decreases in performance.
In contrast, when hypergraph homophily scores are sorted, no clear patterns emerge.

% This result is natural due to the similarity in definitions of $2$-simplicial homophily and the structure of the prediction task.
% Namely, $2$-simplicial homophily is defined based on the proportion of homogenous closed triangles and the proportion of homogenous filled triangles, and the prediction task focuses on whether closed triangles will eventually become filled.
% We note that more general homophily definitions based on structure of the $(k-1)$-skeleton could be used similarly, e.g. in predicting whether an open triangle (three nodes which share two edges) will later become filled.

% Ultimately, the result on link prediction provides a valuable data-driven application for the use of group homophily in networks.
% $2$-simplicial homophily is a useful indicator of whether node label data can assist in the task of prediction.
% \vspace{-0.03in}

\section{Conclusions}
%% more than we studied, what we accomplished. put into past tense.
We proposed a measure for homophily in simplicial complexes, $k$-simplicial homophily, which isolates the  homophily present in group dynamics.
The necessity of such a definition was established on synthetic and empirical data, which indicated that prior notions of homophily for arbitrary hypergraphs can inherit homophilous structure from underlying pairwise interactions and miss the effect of group dynamics.
$k$-simplicial homophily applies to groups of arbitrary size, and we provided experimental and theoretical evidence on triadic interactions that $k$-simplicial homophily provides distinct information from homophily scores on edges.
Moreover, we showed the empirical value of $k$-simplicial homophily, as extreme scores indicate the value of node labels for predicting if group interactions will occur.
These techniques ultimately provide a general approach to isolate group dynamics in simplicial complexes, which we believe will be useful in analyzing group interactions in complex networks more broadly.

\bibliographystyle{plain}
\bibliography{main}

% \appendix
% \section{Dataset Descriptions}
% \input{appendix/data}

% \section{Experimental Details}
% \input{appendix/experiments}
\end{document}